\documentclass[12pt]{article}
\usepackage{epsfig}
\usepackage{multirow}
\usepackage{cite}

\newcommand{\mysection}{\setcounter{equation}{0}\section}

\def\beq{\begin{equation}}
\def\eeq{\end{equation}}
\def\beqa{\begin{eqnarray}}
\def\eeqa{\end{eqnarray}}
 
\newlength{\dinwidth} \newlength{\dinmargin}
\begin{document}

\begin{center}
{\Large \bf Charged Higgs production in association with a top quark at approximate NNLO}
\end{center}
\vspace{2mm}
\begin{center}
{\large Nikolaos Kidonakis}\\
\vspace{2mm}
{\it Department of Physics, Kennesaw State University,\\
Kennesaw, GA 30144, USA}
\end{center}
 
\begin{abstract}
I present approximate next-to-next-to-leading-order (aNNLO) total and differential cross sections
for charged Higgs production in association with a top quark at LHC energies.  
The aNNLO results for the process $bg \rightarrow tH^-$ 
are derived from next-to-next-to-leading-logarithm (NNLL)  
resummation of soft-gluon corrections.  
Scale and parton-distribution uncertainties for the cross sections are shown.
The top-quark transverse-momentum and rapidity distributions are also 
calculated.
\end{abstract}

\mysection{Introduction}

The Higgs sector of the Standard Model and its extensions is a focus of 
particle physics theoretical and experimental programs.  
In two-Higgs-doublet models, such as the minimal supersymmetric standard model,
one Higgs doublet gives mass to the 
up-type fermions while the other to the down-type fermions. 
The ratio of the vacuum expectation values for the two doublets is denoted by 
$\tan \beta$. Among the five Higgs particles in such models there are two charged Higgs bosons, $H^+$ and $H^-$. 

QCD and SUSY-QCD corrections for the associated production of a 
charged Higgs boson and a top quark via the partonic process 
$bg \rightarrow tH^-$ have been calculated through next-to-leading order 
(NLO) in Refs. \cite{BGGS,Zhu,GLXY,Plehn,BHJP,PWRLL,DKSW,FKKSU,DUWZ}. 
The NLO corrections provide a substantial enhancement of the cross section 
and they reduce the scale dependence.

An important class of radiative corrections comes from soft-gluon emission. Near partonic threshold for the production of the $tH^-$ final state these soft-gluon logarithmic corrections are dominant and large; thus, their inclusion is necessary to extend the precision of the theoretical predictions beyond NLO. 

Higher-order soft-gluon corrections for $tH^-$ production were calculated in Refs. \cite{NK04,NKjhep,NKprd73,NKpos,NKprd,NKtop}. The approximate next-to-next-to-leading order (aNNLO) corrections were first derived from next-to-leading-logarithm (NLL) resummation in \cite{NK04,NKjhep} and later improved with next-to-next-to-leading-logarithm (NNLL) resummation in \cite{NKprd}. The aNNLO corrections are significant and they further enhance the cross section.

There have been ongoing active searches for charged Higgs bosons, first at the Tevatron \cite{D0,CDF} and more recently at the LHC \cite{ATLAS,CMS}. 
The increase in the expected theoretical cross section should be taken into account in these searches and in the setting of limits for charged Higgs production in association with a top quark. 

In the next section we provide some theoretical discussion and results for the soft-gluon corrections in $tH^-$ production. In Section 3 we present total cross sections for $tH^-$ production at LHC energies, with various choices of charged Higgs mass as well as $\tan\beta$, noting that the cross sections for ${\bar t}H^+$ production are the same. In Section 4 we present the top-quark transverse-momentum and rapidity distributions in this process. We conclude in Section 5. 

\mysection{Soft-gluon corrections for $tH^-$ production}

We study $tH^-$ production in collisions of protons $A$ and $B$.
The leading-order (LO) diagrams for $bg \rightarrow tH^-$ 
are shown in Fig. {\ref{figLO}.
The hadronic kinematical variables for the process 
$A(p_A)+B(p_B) \rightarrow t(p_t)+H^-(p_{H^-})$
are $S=(p_A+p_B)^2$, $T=(p_A-p_t)^2$, and $U=(p_B-p_t)^2$.
For the partonic reaction $b(p_b)+g(p_g) \rightarrow t(p_t)+H^-(p_{H^-})$,
the partonic kinematical variables are $s=(p_b+p_g)^2$, $t=(p_b-p_t)^2$, 
and $u=(p_g-p_t)^2$, with $p_b=x_1 p_A$ and $p_g=x_2 p_B$. 
We also define the threshold variable $s_4=s+t+u-m_t^2-m_{H^-}^2$, 
where $m_{H^-}$ is the charged Higgs mass and $m_t$ is the top quark mass.    
Note that $s_4$ measures any additional radiation and it vanishes at partonic 
threshold.

\begin{figure}
\begin{center}
\includegraphics[width=10cm]{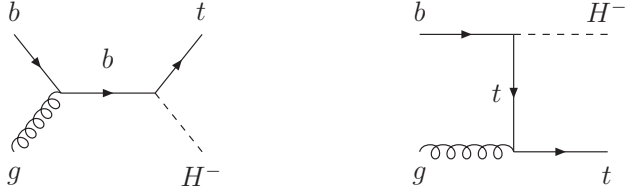}
\caption{Leading-order diagrams for $bg \rightarrow tH^-$.}
\label{figLO}
\end{center}
\end{figure}

The double-differential LO cross section is 
$d^2{\hat\sigma}_{\rm LO}^{bg \rightarrow t H^-}/(dt \; du)
=F_{\rm LO}^{bg \rightarrow t H^-} \delta(s_4)$
where 
\beqa
F_{\rm LO}^{bg \rightarrow t H^-}&=&
\frac{\pi \alpha \alpha_s (m_b^2 \tan^2\beta
+m_t^2 \cot^2\beta)}{12 s^2 m_W^2 \sin^2\theta_w}
\left\{\frac{s+t-m^2_{H^-}}{2s} \right. 
\nonumber \\ && \hspace{-15mm} \left.
{}-\frac{m_t^2(u-m^2_{H^-})+m^2_{H^-}(t-m_t^2)+s(u-m_t^2)}{s(u-m_t^2)}
-\frac{m_t^2(u-m^2_{H^-}-s/2)+su/2}{(u-m_t^2)^2}\right\} ,
\label{FLO}
\eeqa
where $\alpha=e^2/(4\pi)$, $\alpha_s$ is the strong coupling, 
$\theta_w$ is the weak-mixing angle, 
$m_W$ is the $W$ boson mass, and $m_b$ is the $b$-quark mass which is 
taken to be zero everywhere except in the $m_b^2 \tan^2 \beta$ term. 

The perturbative $n$th-order soft-gluon corrections appear as logarithmic plus-distribution enhancements, $[\ln^k(s_4/m_{H^-}^2)/s_4]_+$, with $0 \le k \le 2n-1$. These corrections can be derived from resummation, starting with the factorization properties of the cross section in moment space. We write moments of the partonic cross section ${\hat \sigma}(N)=\int (ds_4/s) \; e^{-N s_4/s} {\hat \sigma}(s_4)$, with $N$ the moment variable. Logarithms of $s_4$ in the physical cross section give rise to logarithms of $N$ in moment space, and those logarithms of $N$ 
exponentiate.

The factorized expression for the moment-space 
partonic cross section in $n=4-\epsilon$ dimensions is
\beq
{\hat \sigma}^{bg \rightarrow tH^-}(N,\epsilon)= 
H^{bg \rightarrow tH^-} \left(\alpha_s(\mu)\right)\; 
S^{bg \rightarrow tH^-} \left(\frac{m_{H^-}}{N \mu},\alpha_s(\mu) \right)
\label{factorsigma}
\eeq 
with $\mu$ the scale. Here the hard function $H^{bg\rightarrow tH^-}$ 
involves contributions from the amplitude of the process and its complex 
conjugate, while $S^{bg\rightarrow tH^-}$ is the soft function for 
noncollinear soft-gluon emission and it represents the coupling of soft gluons
to the partons in the scattering. 

The product of the hard and soft functions 
in Eq. (\ref{factorsigma}) is independent of the gauge and 
the factorization scale, and its evolution results in the exponentiation 
of logarithms of $N$. 
The soft function $S^{bg \rightarrow tH^-}$ requires renormalization and thus  
its $N$-dependence can be resummed via renormalization group evolution. We have 
\beq
S_{\rm bare}^{bg \rightarrow tH^-} =Z_S^{\dagger \, bg \rightarrow tH^-} \, S^{bg \rightarrow tH^-} \, Z_S^{bg \rightarrow tH^-}
\eeq
where $S_{\rm bare}^{bg \rightarrow tH^-}$ is the unrenormalized function,
and $Z_S^{bg \rightarrow tH^-}$ is a renormalization constant.

Thus, $S^{bg \rightarrow tH^-}$ satisfies the renormalization-group equation
\beq
\left(\mu \frac{\partial}{\partial \mu}
+\beta(g_s, \epsilon)\frac{\partial}{\partial g_s}\right)\,S^{bg \rightarrow tH^-}
=- 2 \, S^{bg \rightarrow tH^-} \, \Gamma_S^{bg \rightarrow tH^-} \, .
\eeq
In the above equation $g_s^2=4\pi\alpha_s$; 
$\beta(g_s, \epsilon)=-g_s \epsilon/2 + \beta(g_s)$ 
where $\beta(g_s)$ is the QCD beta function
$\beta(g_s) \equiv \mu d g_s /d \mu$;  
and $\Gamma_S^{bg \rightarrow tH^-} $ is the soft anomalous dimension 
that controls the evolution of the soft function $S^{bg\rightarrow tH^-}$: 
\beq
\Gamma_S^{bg \rightarrow tH^-}=\frac{dZ_S^{bg \rightarrow tH^-}}{d\ln\mu} (Z^{-1})_S^{bg \rightarrow tH^-} \, .
\eeq
Writing $\Gamma_S^{bg \rightarrow tH^-}=(\alpha_s/\pi) \Gamma_S^{(1)}+(\alpha_s/\pi)^2 \Gamma_S^{(2)}+\cdots$, we have \cite{NKprd}
\beq
\Gamma_S^{(1)}=C_F \left[\ln\left(\frac{m_t^2-t}{m_t\sqrt{s}}\right)
-\frac{1}{2}\right] +\frac{C_A}{2} \ln\left(\frac{m_t^2-u}{m_t^2-t}\right)
\label{GS1}
\eeq
and
\beq
\Gamma_S^{(2)}=\left[C_A \left(\frac{67}{36}-\frac{\zeta_2}{2}\right)-\frac{5}{18} n_f\right] \Gamma_S^{(1)} +C_F C_A \frac{(1-\zeta_3)}{4}
\label{GS2}
\eeq
where $C_F=(N_c^2-1)/(2N_c)$ and $C_A=N_c$, with $N_c=3$ the number of colors, 
and $n_f=5$ is the number of light-quark flavors.

The resummed cross section in moment space is derived from the renormalization-group evolution of the hard and soft functions in the factorized cross section, 
Eq. (\ref{factorsigma}), and is given by:
\beqa
{\hat \sigma}^{bg \rightarrow tH^-}_{\rm resummed}(N) &=&
\exp\left[ \sum_{i=b,g} E_i(N_i)\right] \;
H^{bg\rightarrow tH^-}\left(\alpha_s(\sqrt{s})\right)  
\nonumber\\ && 
S^{bg\rightarrow tH^-} \left(\alpha_s\left(\frac{\sqrt{s}}{N'}\right)
\right) \;
\exp \left[\int_{\sqrt{s}}^{\sqrt{s}/N'}
\frac{d\mu}{\mu}\; 2 \, \Gamma_S^{bg\rightarrow tH^-}
\left(\alpha_s(\mu)\right)\right] 
\label{resumsigma}
\eeqa
where the first exponent resums soft and collinear radiation from the incoming bottom quark and gluon, and the second exponent resums noncollinear soft-gluon emission (see \cite{NKprd} for more details).

The resummed cross section in moment space can be expanded in powers of the strong coupling and inverted back to momentum space, thus providing approximate results for the higher-order corrections from soft-gluon emission. The aNNLO soft-gluon corrections in the double-differential partonic cross section, $d^2{\hat \sigma}/(dt \,du)$, can be written as  
\beq 
\frac{d^2{\hat \sigma}_{\rm aNNLO}^{(2) \, bg \rightarrow tH^-}}{dt \, du}=F_{\rm LO}^{bg \rightarrow tH^-} \frac{\alpha_s^2}{\pi^2} 
\sum_{k=0}^3 C_k^{(2)} \left[\frac{\ln^k(s_4/m_{H^-}^2)}{s_4}\right]_+
\label{dtdu}
\eeq
where the superscript ``(2)'' in $\hat \sigma$ and $C_k^{(2)}$ indicates that these are second-order corrections in the strong coupling.
The leading aNNLO coefficient, $C_3^{(2)}$, depends only on color factors:  
$C_3^{(2)}=2(C_F+C_A)^2$.

The subleading coefficients $C_2^{(2)}$, $C_1^{(2)}$, and $C_0^{(2)}$
are in general functions of $s$, $t$, $u$, $m_{H^-}$, $m_t$, and the 
factorization scale $\mu_F$, and - in the case of $C_1^{(2)}$ and $C_0^{(2)}$ - 
also the renormalization scale $\mu_R$. These coefficients have been determined
from one-loop \cite{NK04,NKjhep} and two-loop \cite{NKprd} calculations.
The next-to-leading coefficient is 
\beqa
C_2^{(2)}&=&(C_F+C_A) \left\{3C_F\left[2\ln\left(\frac{m_t^2-t}{m_t \sqrt{s}}\right)-2\ln\left(\frac{m_{H^-}^2-u}{m_{H^-}^2}\right)-1\right] \right.
\nonumber \\ && \left. 
{}-3 C_A \left[\ln\left(\frac{m_t^2-t}{m_t^2-u}\right)
+2 \ln\left(\frac{m_{H^-}^2-t}{m_{H^-}^2}\right)\right]
-3(C_F+C_A) \ln\left(\frac{\mu_F^2}{s}\right)-\frac{\beta_0}{2} \right\}
\eeqa
where $\beta_0=(11C_A-2n_f)/3$.
The expressions for $C_1^{(2)}$ and $C_0^{(2)}$ are much longer \cite{NKprd}.
With NLL resummation \cite{NK04,NKjhep} we can 
calculate all aNNLO coefficients except $C_0^{(2)}$, which can only be fully 
determined by NNLL resummation \cite{NKprd}. 
The one-loop soft anomalous dimension, $\Gamma_S^{(1)}$,  
contributes to all subleading coefficients, while the two-loop soft anomalous 
dimension, $\Gamma_S^{(2)}$, contributes to $C_0^{(2)}$.

The hadronic differential cross section can be calculated via a convolution of the partonic cross section with parton distribution functions (pdf). 
The double-differential cross section with respect to the top-quark transverse momentum, $p_T$, and rapidity, $Y$, is given by
\beq
\frac{d\sigma}{dp_T \, dY}=
2 \, p_T \int_{x_{2{\rm min}}}^1 dx_2 
\int_0^{s_{4 {\rm max}}} ds_4 \, 
\frac{x_1 x_2 \, S}{x_2 S+T_1} \,
\phi(x_1) \, \phi(x_2) \, 
\frac{d^2{\hat\sigma}^{bg \rightarrow tH^-}}{dt \, du}
\eeq
where $T_1=T-m_t^2=-\sqrt{S} \, (m_t^2+p_T^2)^{1/2} \, e^{-Y}$, $U_1=U-m_t^2=-\sqrt{S} \, (m_t^2+p_T^2)^{1/2} \, e^{Y}$, $x_{2{\rm min}}=(m_{H^-}^2-T)/(S+U_1)$,
$s_{4 {\rm max}}=x_2(S+U_1)+T-m_{H^-}^2$, 
$x_1=(s_4-m_t^2+m_{H^-}^2-x_2U_1)/(x_2 S+T_1)$; and  $\phi$ denotes the pdf.
The transverse-momentum and rapidity distributions as well as the total cross section can be obtained by appropriate integrations over this double-differential cross section.

\mysection{$tH^-$ total cross sections}

\begin{figure}
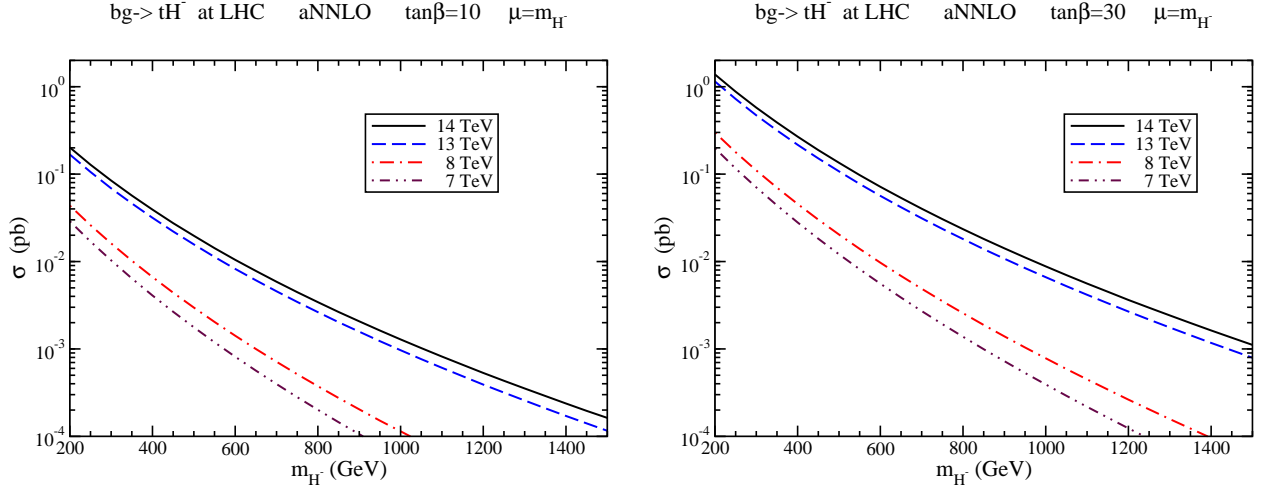

\begin{center}
\includegraphics[width=8cm]{chiggstn10plot.eps}
\hspace{3mm}
\includegraphics[width=8cm]{chiggstn30plot.eps}
\caption{Total aNNLO cross sections for $tH^-$ production 
at 7, 8, 13, and 14 TeV LHC energy with (left) $\tan \beta=10$ and 
(right) $\tan \beta=30$.}
\label{chiggsplot}
\end{center}
\end{figure}

\begin{figure}
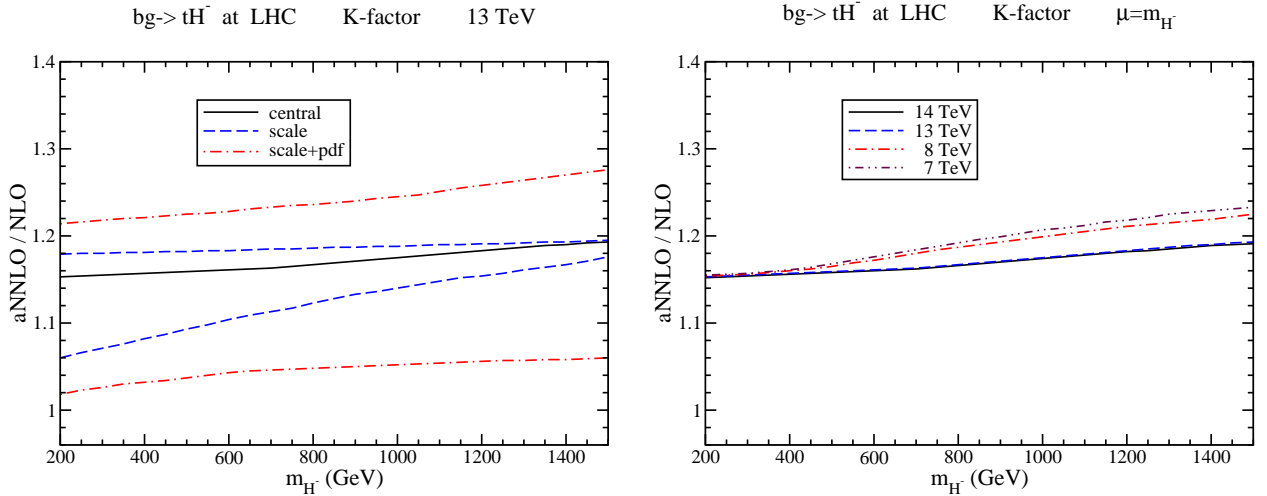

\begin{center}
\includegraphics[width=8cm]{Kchiggs13lhcplot.eps}
\hspace{3mm}
\includegraphics[width=8cm]{Kchiggslhcplot.eps}
\caption{aNNLO/NLO $K$-factors for the total cross section for $tH^-$ 
production at (left) 13 TeV LHC energy and (right) 7, 8, 13, and 14 TeV 
LHC energies.}
\label{Kchiggsplot}
\end{center}
\end{figure}

We continue with results for the total cross sections for charged Higgs production in association with a top quark via the partonic process $bg \rightarrow tH^-$ at LHC energies. The results we present are for $tH^-$ production; the cross sections for ${\bar t} H^+$} production are identical. We use the MMHT2014 \cite{MMHT} NNLO pdf for all our numerical results in this paper.

In Fig. \ref{chiggsplot} we plot the total cross sections at aNNLO for $tH^-$ 
production as functions of charged Higgs mass at 7, 8, 13, and 14 TeV energies 
at the LHC. The left plot shows the results with $\tan\beta=10$ while the 
right plot shows the corresponding results with $\tan\beta=30$. The scales are set equal to the charged Higgs mass. Since the $\tan \beta$ dependence is simply given by the $(m_b^2 \tan^2\beta+m_t^2 \cot^2\beta)$ factor in Eq. (\ref{FLO}), one can easily rescale the results to any desired $\tan\beta$ value. We note that the cross sections decrease over three orders of magnitude as the charged Higgs mass is increased from 200 GeV to 1500 GeV. 

\begin{table}[htb]
\begin{center}
\begin{tabular}{|c|c|c|c|c|} \hline
\multicolumn{5}{|c|}{aNNLO  $tH^-$ cross section with $\tan\beta=30$ at LHC (fb)} \\ \hline
$m_{H^-}$ (GeV) & 7 TeV & 8 TeV & 13 TeV & 14 TeV \\ \hline
200  & 198   & 299   & 1148 & 1382 \\ \hline 
400  & 28.0  & 45.5  & 217  & 269  \\ \hline 
600  & 5.58  & 9.70  & 56.4 & 71.7 \\ \hline 
800  & 1.38  & 2.57  & 18.1 & 23.5 \\ \hline
1000 & 0.393 & 0.782 & 6.62 & 8.81 \\ \hline
\end{tabular}
\caption[]{The aNNLO $tH^-$ production cross section in fb    
in $pp$ collisions at the LHC with $\sqrt{S}=7$, 8, 13, and 14 TeV. We set 
$\tan\beta=30$ and $\mu=m_{H^-}$, and we use the MMHT2014 NNLO pdf \cite{MMHT}.}
\label{table1}
\end{center}
\end{table}

In Table \ref{table1} we provide some numbers for the total cross sections for selected values of charged Higgs mass at LHC energies of 7, 8, 13, and 14 TeV.
The values are for $\tan\beta=30$ and scale $\mu=m_{H^-}$.

In Fig. \ref{Kchiggsplot} we plot the aNNLO/NLO ratios, i.e. $K$ factors, for the total cross sections for $tH^-$ production as functions of charged Higgs mass. The $\tan\beta$ dependence of course cancels out in $K$ factors. The left plot displays results at 13 TeV LHC energy. The line labeled as ``central'' is with the scales set equal to the charged Higgs mass. The lines labeled as ``scale'' show the variation when the scales in the aNNLO cross section are varied by a factor of two. The lines labeled as ``scale + pdf'' show the total uncertainty including scale variation plus uncertainties from the parton distribution functions as given by \cite{MMHT}. We see that the $K$ factors are sizable, indicating significant contributions of 15\% to 20\% - depending on the charged Higgs mass - from the central aNNLO corrections. The right plot shows the central $K$ factors at 7, 8, 13, and 14 TeV LHC energies. The $K$ factors are larger at the smaller LHC energies, since then the process is closer to partonic threshold.

\mysection{Top-quark $p_T$ and rapidity distributions}

\begin{figure}
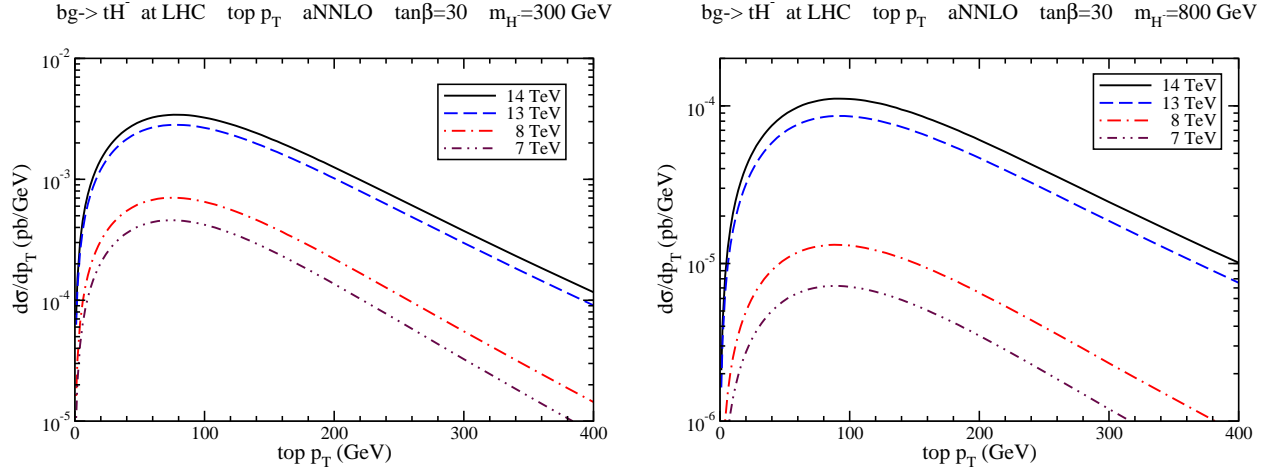

\begin{center}
\includegraphics[width=8cm]{pttopchiggs300tn30plot.eps}
\hspace{3mm}
\includegraphics[width=8cm]{pttopchiggs800tn30plot.eps}
\caption{aNNLO top-quark $p_T$ distributions, $d\sigma/dp_T$, in $tH^-$ 
production with $\tan \beta=30$ at 7, 8, 13, and 14 TeV 
LHC energy with (left) $m_{H^-}=300$ GeV and (right) $m_{H^-}=800$ GeV. }
\label{pttopchiggstn30plot}
\end{center}
\end{figure}

\begin{figure}
\begin{center}
\includegraphics[width=8cm]{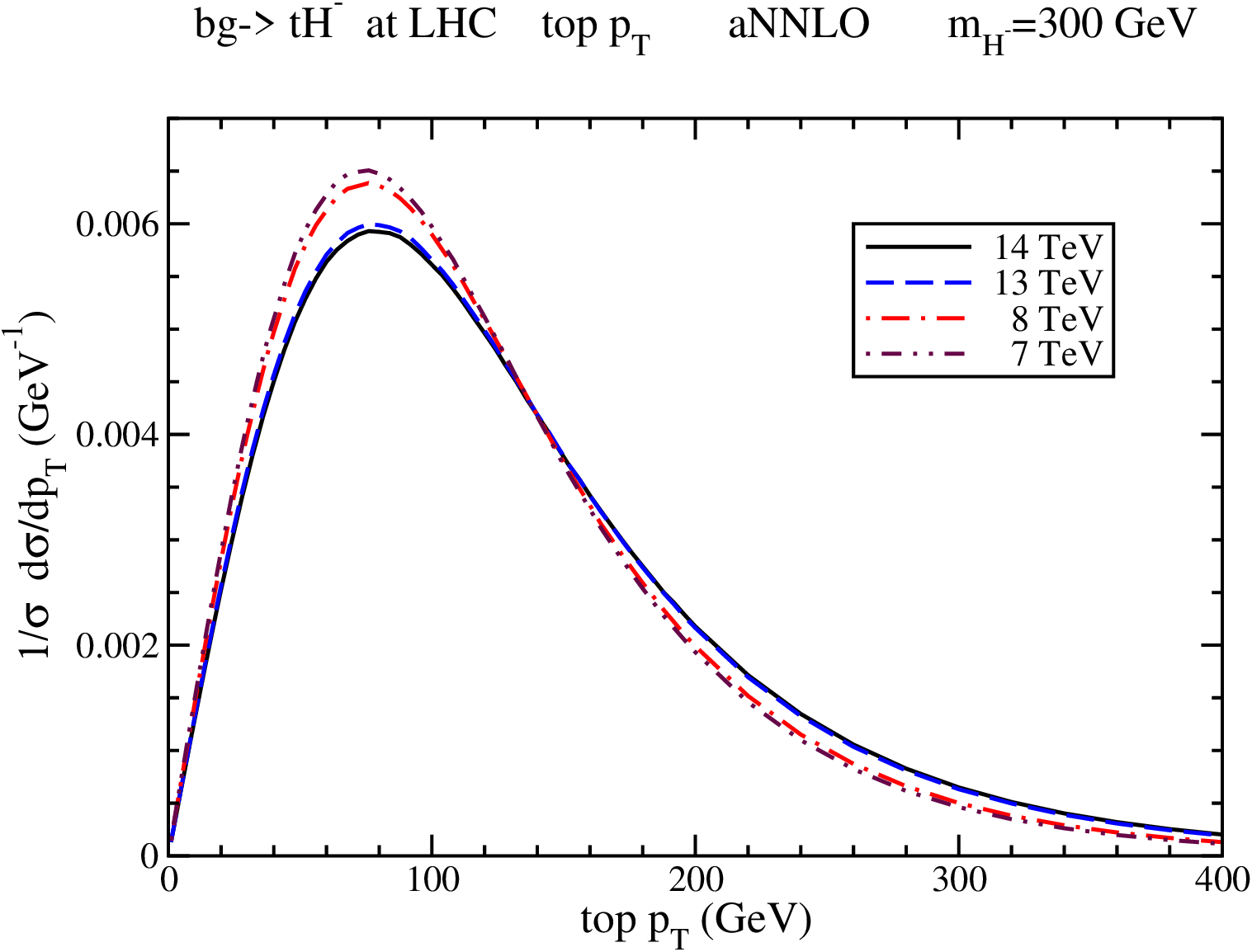}
\hspace{3mm}
\includegraphics[width=8cm]{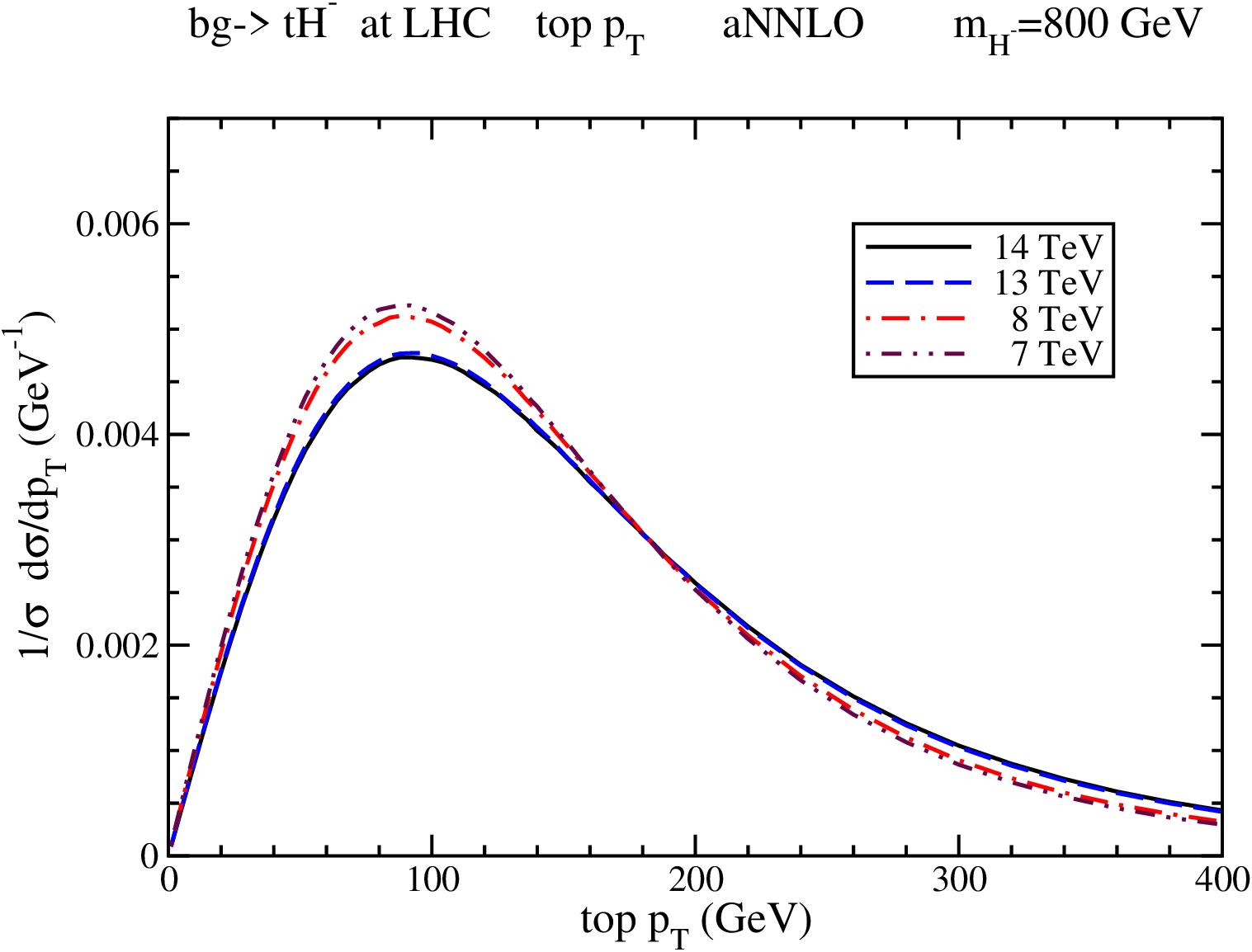}
\caption{aNNLO top-quark normalized $p_T$ distributions, $(1/\sigma) d\sigma/dp_T$, in $tH^-$ production at 7, 8, 13, and 14 TeV LHC energy with (left) $m_{H^-}=300$ GeV and (right) $m_{H^-}=800$ GeV.}
\label{ptnormtopchiggsplot}
\end{center}
\end{figure}

\begin{figure}
\begin{center}
\includegraphics[width=8cm]{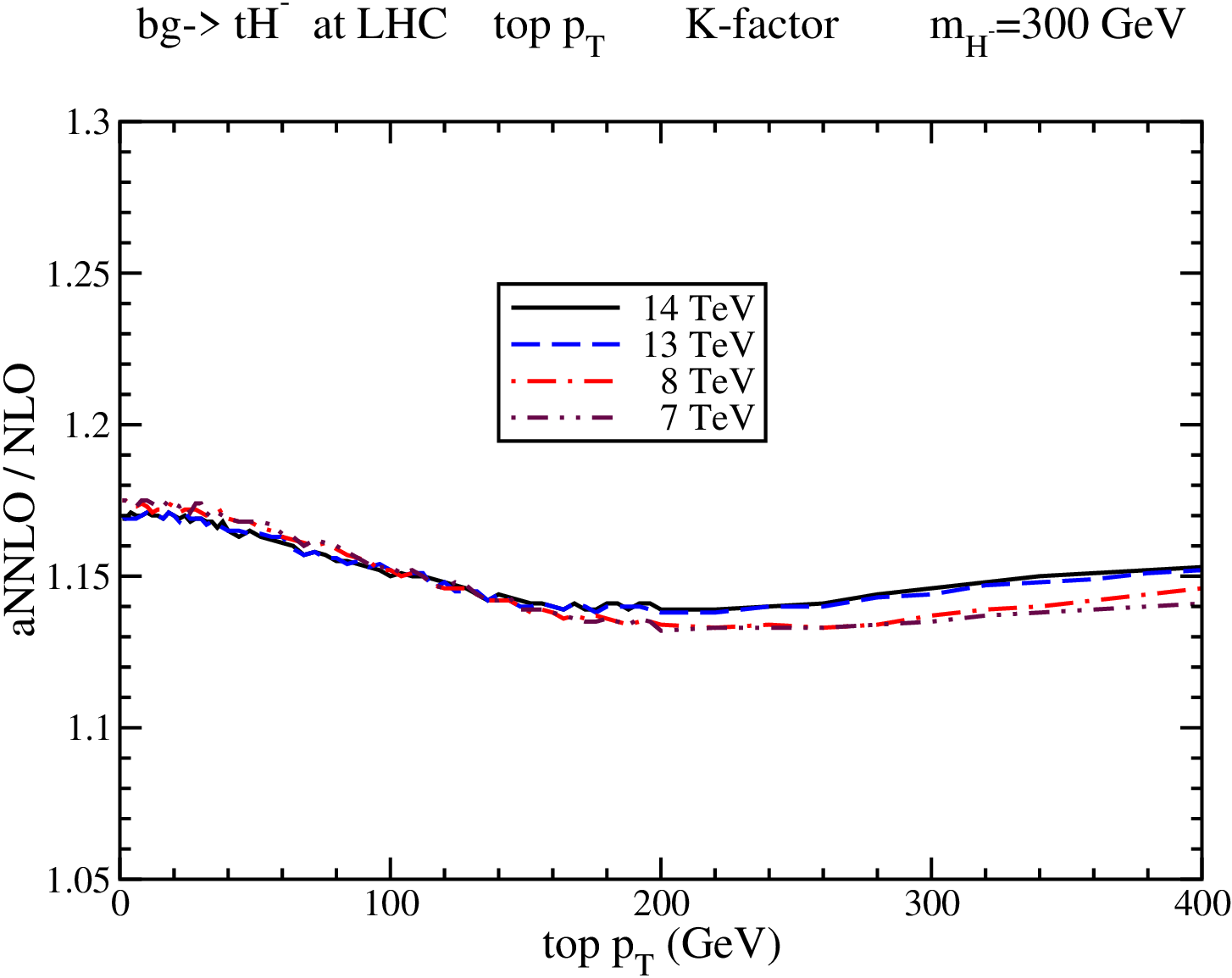}
\hspace{3mm}
\includegraphics[width=8cm]{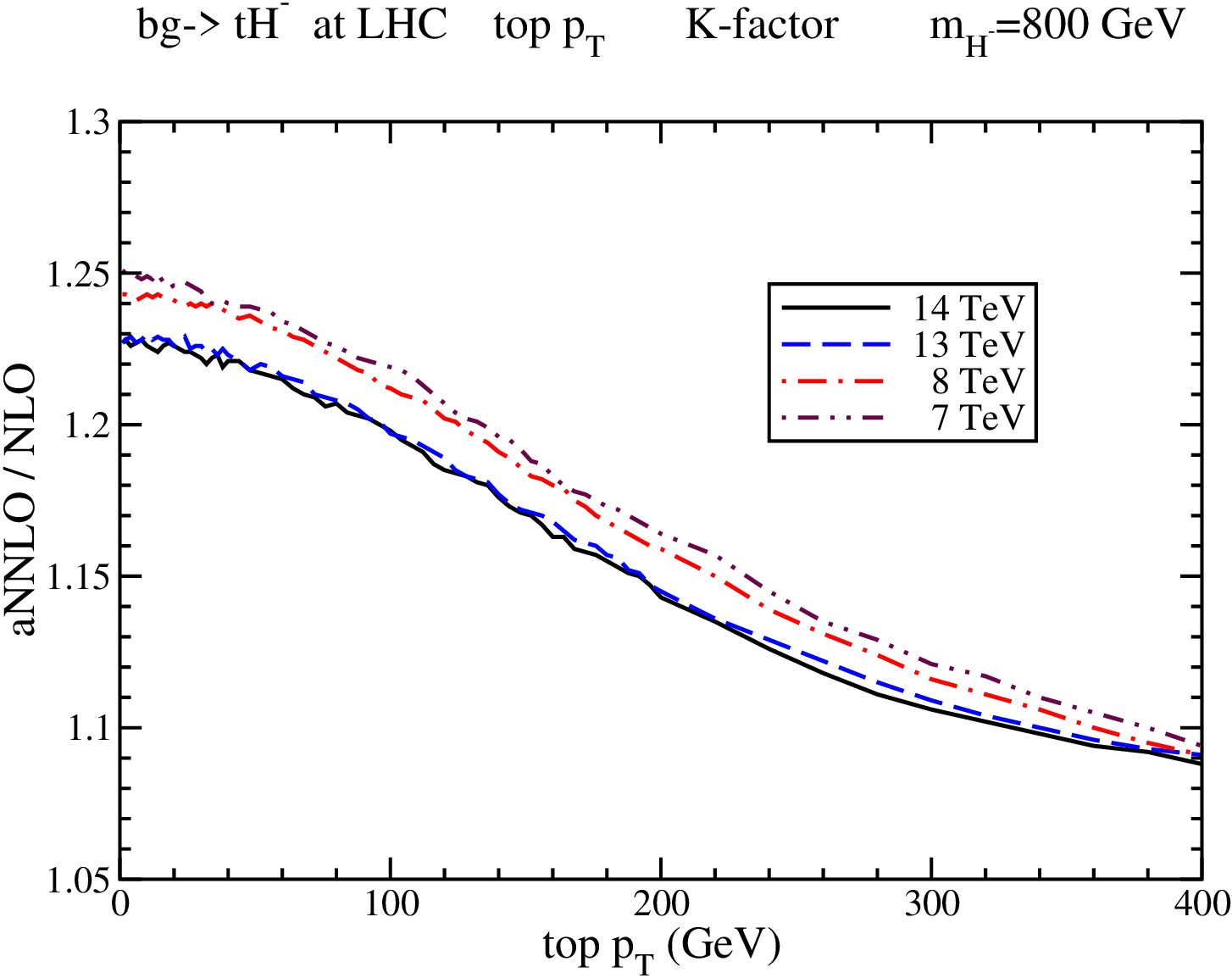}
\caption{The aNNLO/NLO $K$-factors for the top-quark $p_T$ distributions, $d\sigma/dp_T$, in $tH^-$ production at 7, 8, 13, and 14 TeV LHC energy with (left) $m_{H^-}=300$ GeV and (right) $m_{H^-}=800$ GeV.}
\label{Kpttopchiggsplot}
\end{center}
\end{figure}

We continue with a presentation of differential distributions in $tH^-$ production, in particular the top-quark transverse-momentum and rapidity distributions.

In Fig. \ref{pttopchiggstn30plot} we plot the top-quark transverse-momentum 
distributions, $d\sigma/dp_T$, at aNNLO with $\tan\beta=30$ for 7, 8, 13, and 
14 TeV LHC energies. The left plot uses $m_{H^-}=300$ GeV and the right plot 
uses $m_{H^-}=800$ GeV. One can derive results for any value of $\tan\beta$ with 
appropriate rescaling. For $m_{H^-}=300$ GeV the distributions peak at $p_T$ values of around 75 GeV and drop by two orders of magnitude at a $p_T$ of 400 GeV. For $m_{H^-}=800$ GeV the distributions are much smaller due to the very heavy final state.

In Fig. \ref{ptnormtopchiggsplot} we plot the normalized top-quark transverse-momentum distributions,  $(1/\sigma) d\sigma/dp_T$, at aNNLO for LHC energies. The left plot is with $m_{H^-}=300$ GeV and the right plot is with $m_{H^-}=800$ GeV. The use of normalized distributions removes the $\tan\beta$ dependence and minimizes the dependence on the pdf. The dependence on the charged Higgs mass is also milder for the normalized distributions. We note that at lower LHC energies the normalized top $p_T$ distribution is somewhat smaller at large $p_T$ than it is at higher energies, as expected.

Figure \ref{Kpttopchiggsplot} shows the aNNLO/NLO ratios, i.e. $K$ factors, for the top-quark transverse-momentum distributions at LHC energies. The left plot has $m_{H^-}=300$ GeV and the right plot has $m_{H^-}=800$ GeV. The aNNLO contributions are clearly seen to be quite significant, around 15\% for most $p_T $ values when $m_{H^-}=300$ GeV. Again, the $\tan\beta$ dependence cancels out in the $K$ factors. 

\begin{figure}
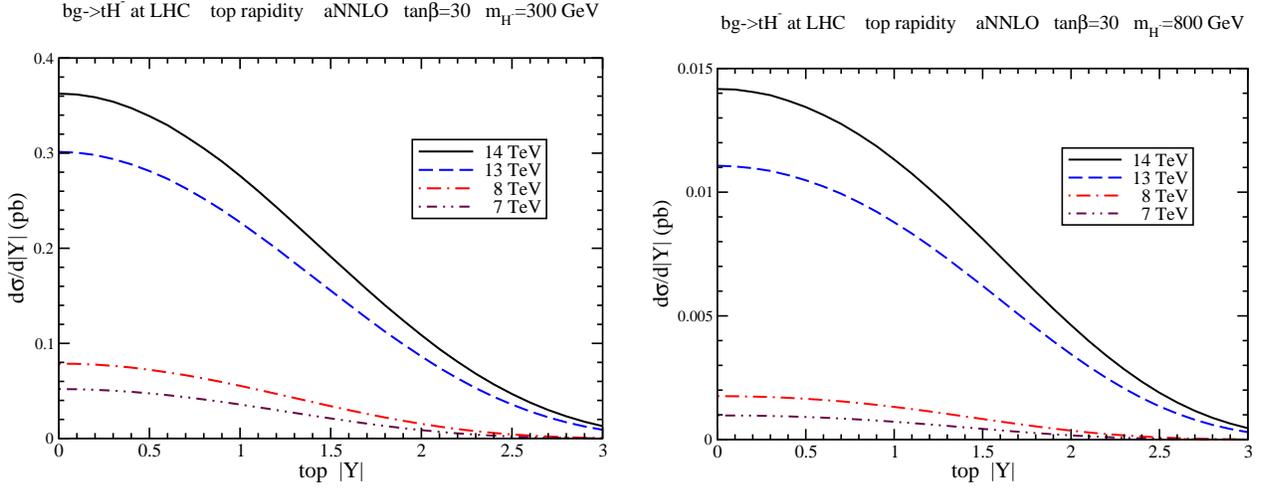

\begin{center}
\includegraphics[width=8cm]{yabstopchiggs300tn30plot.eps}
\hspace{3mm}
\includegraphics[width=8cm]{yabstopchiggs800tn30plot.eps}
\caption{aNNLO top-quark rapidity distributions, $d\sigma/d|Y|$, in $tH^-$ 
production with $\tan \beta=30$ at 7, 8, 13, and 14 TeV 
LHC energy with (left) $m_{H^-}=300$ GeV and (right) $m_{H^-}=800$ GeV.}
\label{yabstopchiggstn30plot}
\end{center}
\end{figure}

\begin{figure}
\begin{center}
\includegraphics[width=8cm]{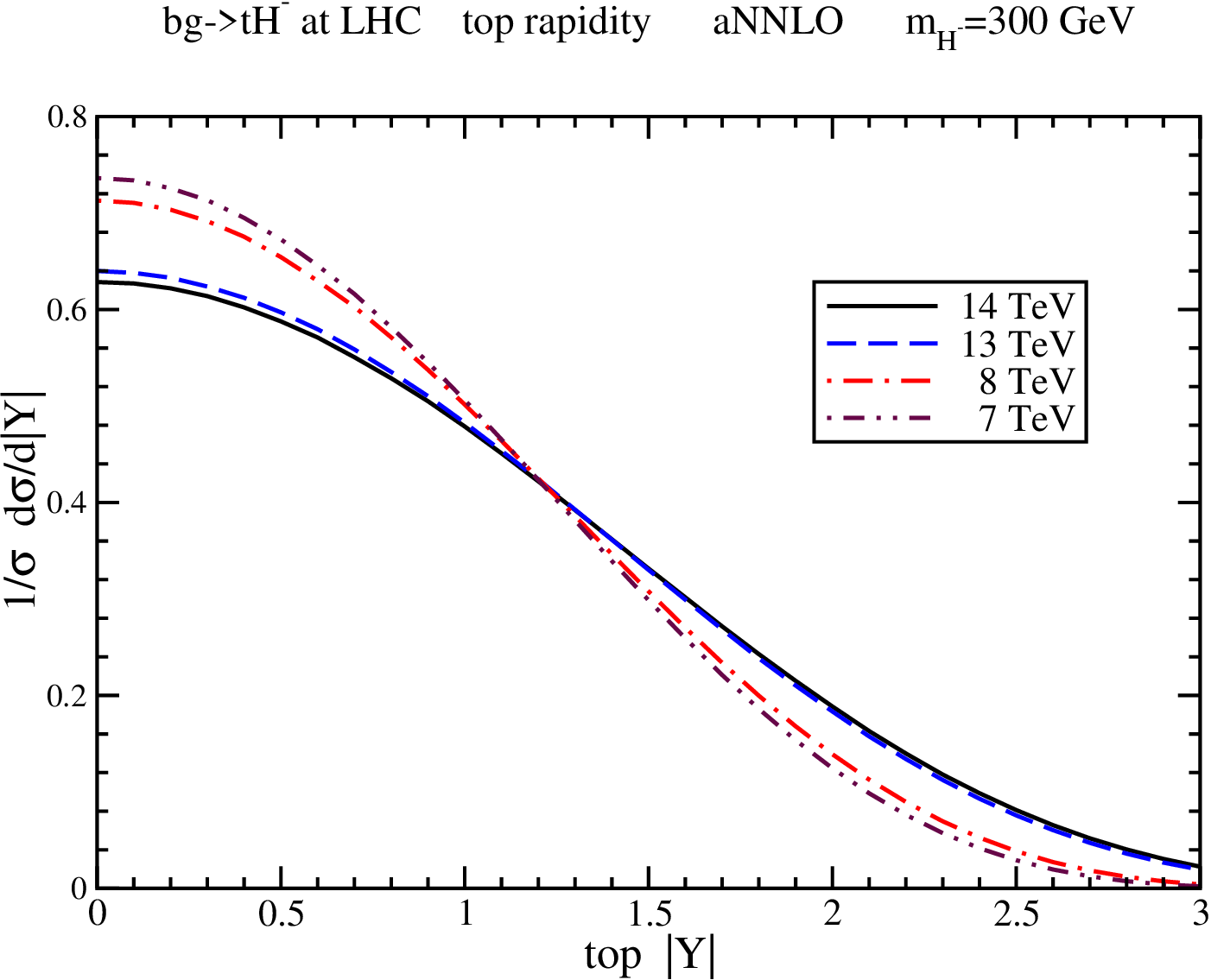}
\hspace{3mm}
\includegraphics[width=8cm]{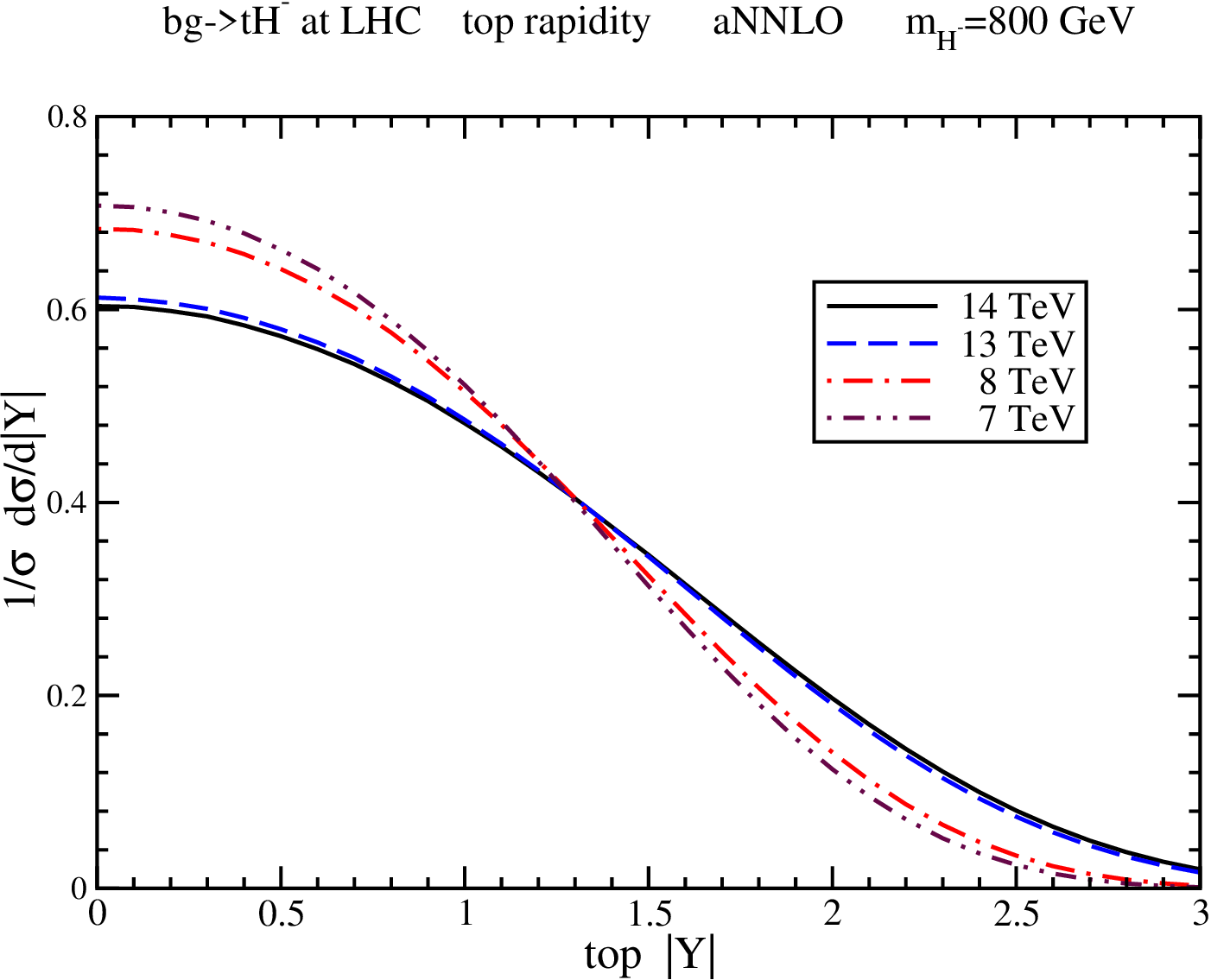}
\caption{aNNLO top-quark normalized rapidity distributions, $(1/\sigma) d\sigma/d|Y|$, in $tH^-$ production at 7, 8, 13, and 14 TeV LHC energy with (left) $m_{H^-}=300$ GeV and (right) $m_{H^-}=800$ GeV.}
\label{yabsnormtopchiggsplot}
\end{center}
\end{figure}

\begin{figure}
\begin{center}
\includegraphics[width=8cm]{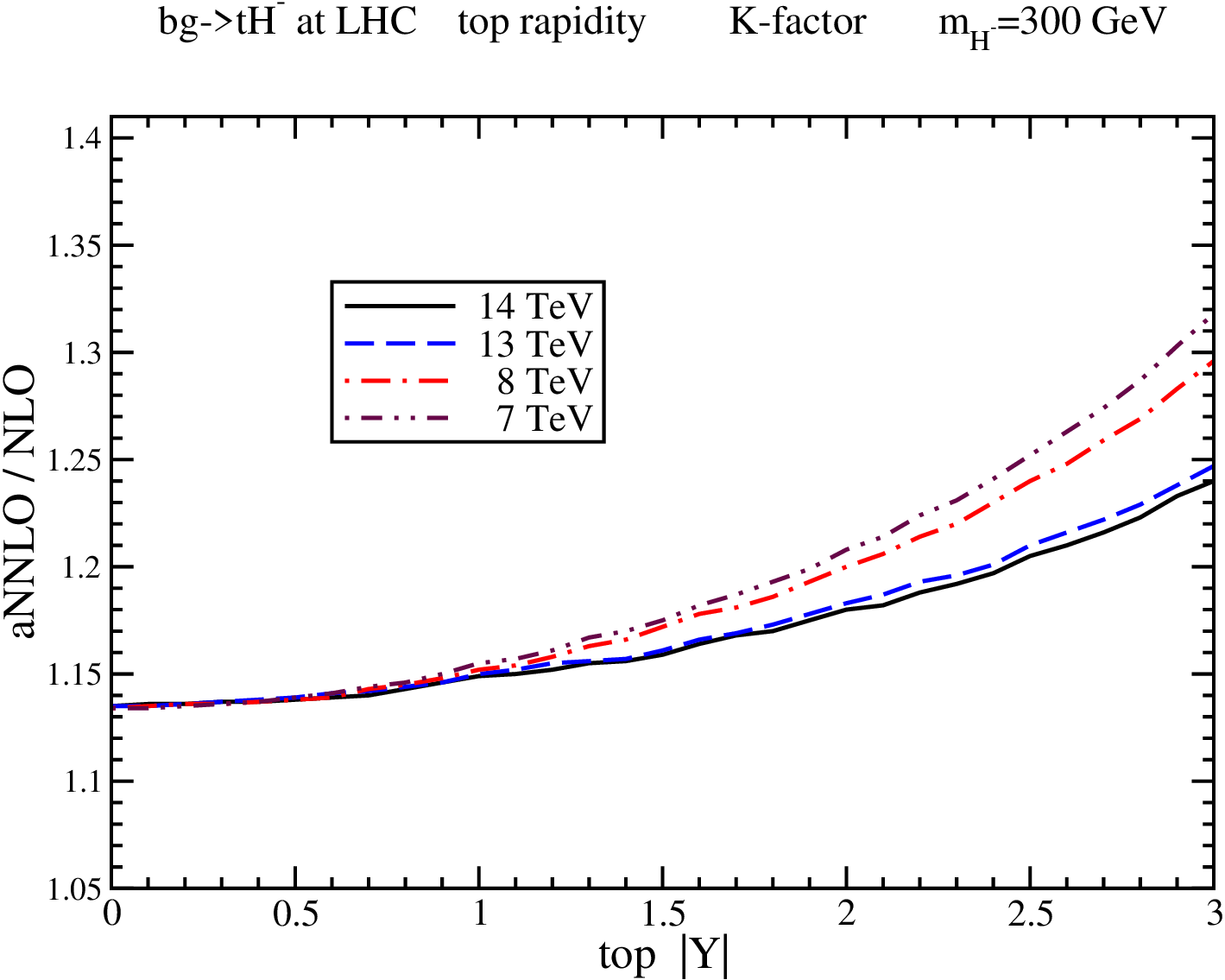}
\hspace{3mm}
\includegraphics[width=8cm]{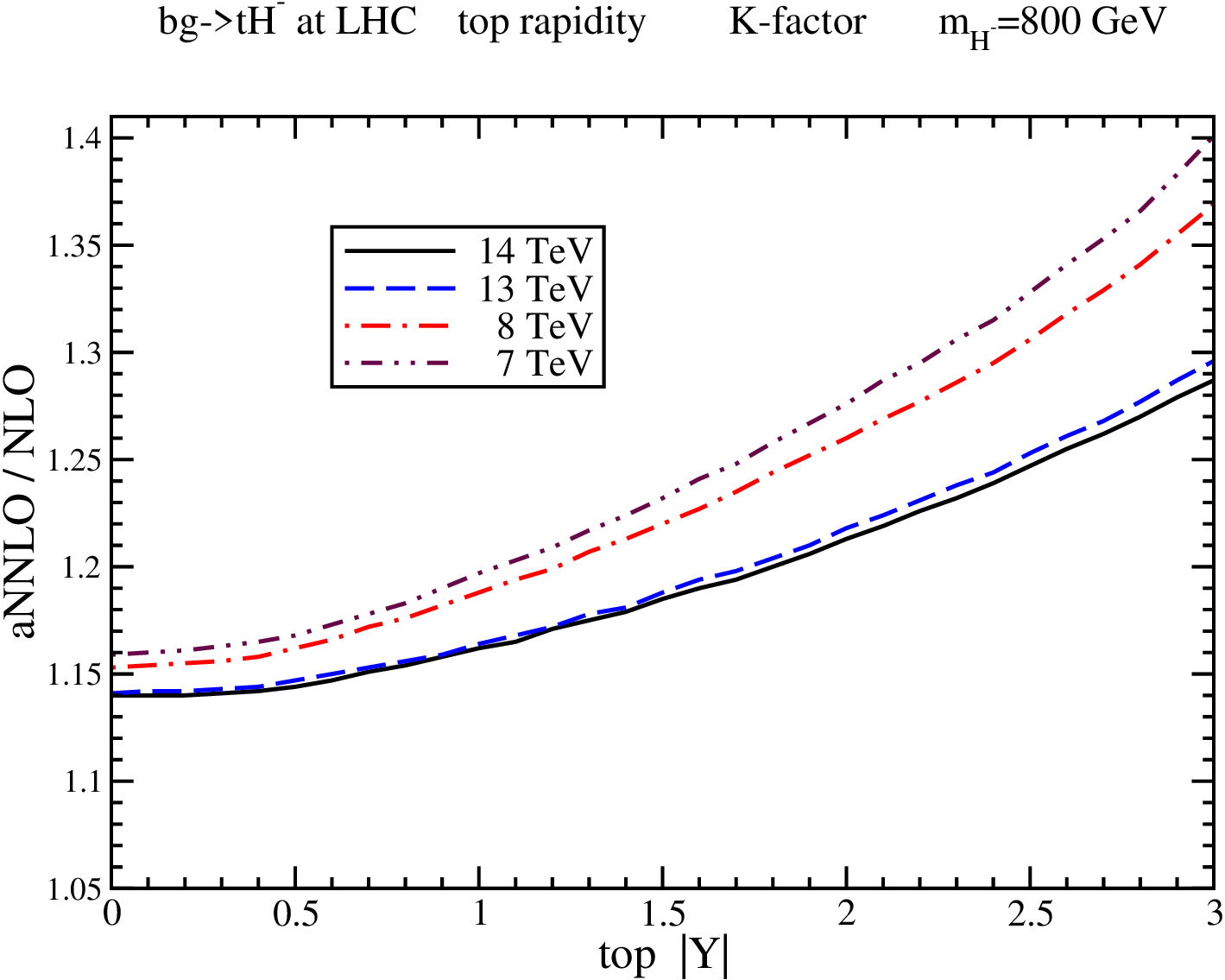}
\caption{The aNNLO/NLO $K$-factors for the top-quark rapidity distributions, $d\sigma/d|Y|$, in $tH^-$ production at 7, 8, 13, and 14 TeV LHC energy with (left) $m_{H^-}=300$ GeV and (right) $m_{H^-}=800$ GeV.}
\label{Kyabstopchiggsplot}
\end{center}
\end{figure}

In Fig. \ref{yabstopchiggstn30plot} we plot the top-quark rapidity distributions, $d\sigma/d|Y|$, at aNNLO with $\tan\beta=30$ for 7, 8, 13, and 14 TeV LHC energies. The left plot uses $m_{H^-}=300$ GeV and the right plot uses $m_{H^-}=800$ GeV. The distributions at 800 GeV mass are of course much smaller. As before, one can derive results for any value of $\tan\beta$ with appropriate rescaling.

In Fig. \ref{yabsnormtopchiggsplot} we plot the normalized top-quark rapidity
distributions,  $(1/\sigma) d\sigma/d|Y|$, at aNNLO for LHC energies. The left plot is with $m_{H^-}=300$ GeV and the right plot is with $m_{H^-}=800$ GeV. We note that at lower energies the normalized top rapidity distribution is smaller at large absolute values of $|Y|$ than it is at higher energies, as expected.

Figure \ref{Kyabstopchiggsplot} shows the aNNLO/NLO ratios, i.e. $K$ factors, for the top-quark rapidity distributions at LHC energies. The left plot has 
$m_{H^-}=300$ GeV and the right plot has $m_{H^-}=800$ GeV. The aNNLO contributions are quite significant, especially at large rapidity. At 7 TeV LHC energy and 800 GeV mass, the aNNLO corrections are 40\% at $|Y|=3$.

\mysection{Conclusions}

I have presented aNNLO results for charged Higgs production in association with a top quark via the partonic process $b g \rightarrow tH^-$. Total cross sections and top-quark $p_T$ and rapidity distributions have been calculated at LHC energies. The NNLO soft-gluon contributions are important and provide significant enhancements of the total cross sections. Uncertainties due to scale variations and due to the parton distribution functions have also been presented.

The top-quark differential distributions in transverse momentum and rapidity also receive significant enhancements from soft-gluon contributions. The corrections are particularly large at large rapidity values. 

The aNNLO corrections need to be included in theoretical predictions for total cross sections and differential distributions in order to provide more precision in the search for charged Higgs bosons.

\mysection*{Acknowledgements}
This material is based upon work supported by the National Science Foundation 
under Grant No. PHY 1519606.

\end{document}